\documentclass[sigconf]{acmart}

\makeatletter
\def\mdseries@tt{m}
\makeatother
\usepackage[frozencache=true]{minted}

\renewcommand\footnotetextcopyrightpermission[1]{}
\AtBeginDocument{%
  \providecommand\BibTeX{{%
    \normalfont B\kern-0.5em{\scshape i\kern-0.25em b}\kern-0.8em\TeX}}}

\usepackage{balance} 

\usepackage{subcaption}
\usepackage{xcolor}
\usepackage{hyperref}
\usepackage{amsmath}
\usepackage{amsfonts}
\usepackage{booktabs} 
\usepackage{multirow}
\usepackage{rotating}
\usepackage{listings}
\usepackage{wrapfig}
\usepackage[export]{adjustbox}

\usepackage{makecell,cellspace}
\usepackage{fontawesome}
\newcommand{\thumbsup}{\faThumbsOUp}
\newcommand{\thumbsdown}{\faThumbsDown}

\fvset{linenos, fontsize=\small, frame=lines, numbersep=1pt, xleftmargin=.75em}
\DeclareCaptionSubType{listing} 

\usepackage{bm}
\newcommand{\matr}[1]{\bm{#1}}
\renewcommand\vec{\mathbf} 

\renewcommand{\paragraph}[1]{{\noindent\textbf{\textit{#1}}~~~}}

\newcommand{\figref}[1]{Fig.~\ref{fig:#1}}
\newcommand{\tabref}[1]{Tab.~\ref{tab:#1}}
\newcommand{\lstref}[1]{Lst.~\ref{lst:#1}}
\newcommand{\equref}[1]{Eq.~\ref{eq:#1}}
\newcommand{\secref}[1]{Sec.~\ref{sec:#1}}

\usepackage{mathtools} 
\DeclarePairedDelimiter{\ceil}{\lceil}{\rceil}
\DeclarePairedDelimiter{\floor}{\lfloor}{\rfloor}

\usepackage[per-mode=symbol]{siunitx}
\DeclareSIUnit \bit {bit}
\DeclareSIUnit \byte {B}
\DeclareSIUnit \Byte {Byte}
\DeclareSIUnit \cycle {cycle}
\DeclareSIUnit \cycles {cycles}
\DeclareSIUnit \hz {Hz}
\DeclareSIUnit \op {Op}
\DeclareSIUnit \operand {operand}
\DeclareSIUnit \operands {operands}
\DeclareSIUnit \transfer {T}
\DeclareSIUnit \cell {cell}

\usepackage[colorinlistoftodos, prependcaption, textsize=tiny]{todonotes}
\usepackage{xargs}
\newcommandx{\unsure}[2][1=]{\todo[linecolor=red, backgroundcolor=red!25, bordercolor=red, #1]{#2}}
\newcommandx{\change}[2][1=]{\todo[linecolor=blue, backgroundcolor=blue!25, bordercolor=blue, #1]{#2}}
\newcommandx{\info}[2][1=]{\todo[linecolor=OliveGreen, backgroundcolor=OliveGreen!25, bordercolor=OliveGreen, #1]{#2}}
\newcommandx{\improvement}[2][1=]{\todo[linecolor=Plum, backgroundcolor=Plum!25, bordercolor=Plum, #1]{#2}}

\title[Flexible Communication Avoiding Matrix Multiplication on FPGA with High-Level
Synthesis]{Flexible Communication Avoiding Matrix Multiplication \\on FPGA with
High-Level Synthesis}

\author{Johannes de Fine Licht}
\affiliation{%
  \institution{ETH Zurich}
}
\email{definelicht@inf.ethz.ch}

\author{Grzegorz Kwasniewski}
\affiliation{%
  \institution{ETH Zurich}
}
\email{gkwasnie@inf.ethz.ch}

\author{Torsten Hoefler}
\affiliation{%
  \institution{ETH Zurich}
}
\email{htor@inf.ethz.ch}

\begin{abstract}
Data movement is the dominating factor affecting performance and energy in
modern computing systems.
Consequently, many algorithms have been developed to minimize the number of I/O
operations for common computing patterns.
Matrix multiplication is no exception, and lower bounds have been proven and
implemented both for shared and distributed memory systems.
Reconfigurable hardware platforms are a lucrative target for I/O minimizing
algorithms, as they offer full control of memory accesses to the programmer.
While bounds developed in the context of fixed architectures still apply to
these platforms, the spatially distributed nature of their computational and
memory resources requires a decentralized approach to optimize algorithms for
maximum hardware utilization. 
We present a model to optimize matrix multiplication for FPGA platforms,
simultaneously targeting maximum performance and minimum off-chip data movement,
within constraints set by the hardware.
We map the model to a concrete architecture using a high-level synthesis tool,
maintaining a high level of abstraction, allowing us to support arbitrary data
types, and enables maintainability and portability across FPGA devices.
Kernels generated from our architecture are shown to offer competitive
performance in practice, scaling with both compute and memory resources.
We offer our design as an open source
project\footnote{\url{https://github.com/spcl/gemm_hls}
(\href{https://doi.org/10.5281/zenodo.3559536}{10.5281/zenodo.3559536})} to
encourage the open development of linear algebra and I/O minimizing algorithms
on reconfigurable hardware platforms.
\end{abstract}

\begin{document}
\fancyhead{}

\maketitle

\begin{figure}[t]
  \centering
  \includegraphics[width=.95\columnwidth]{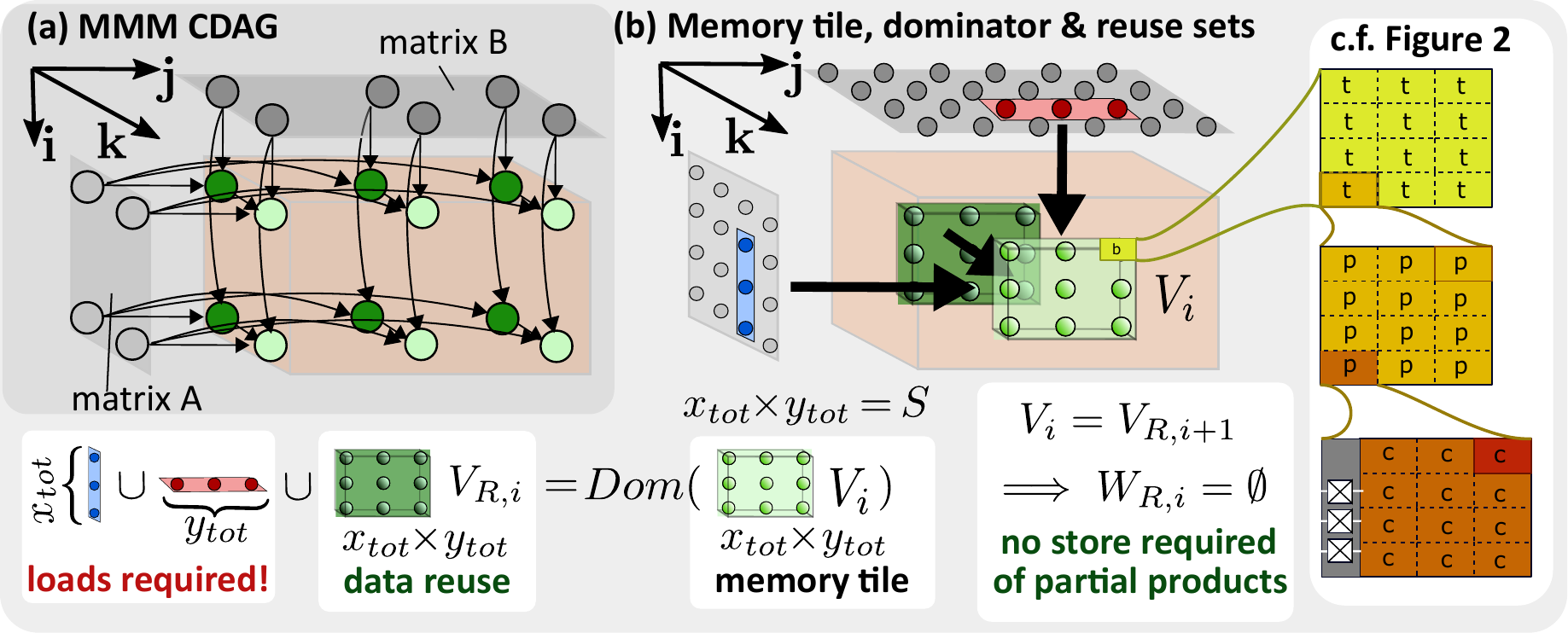}
  \vspace{-0.5em}
  \caption{(a) MMM CDAG\@, and (b) subcomputation $V_i$.}
  \vspace{-2em}
  \label{fig:mmm_cdag}
\end{figure}

\begin{figure*}[t]
	\centering
	\includegraphics[width=0.95 \textwidth]{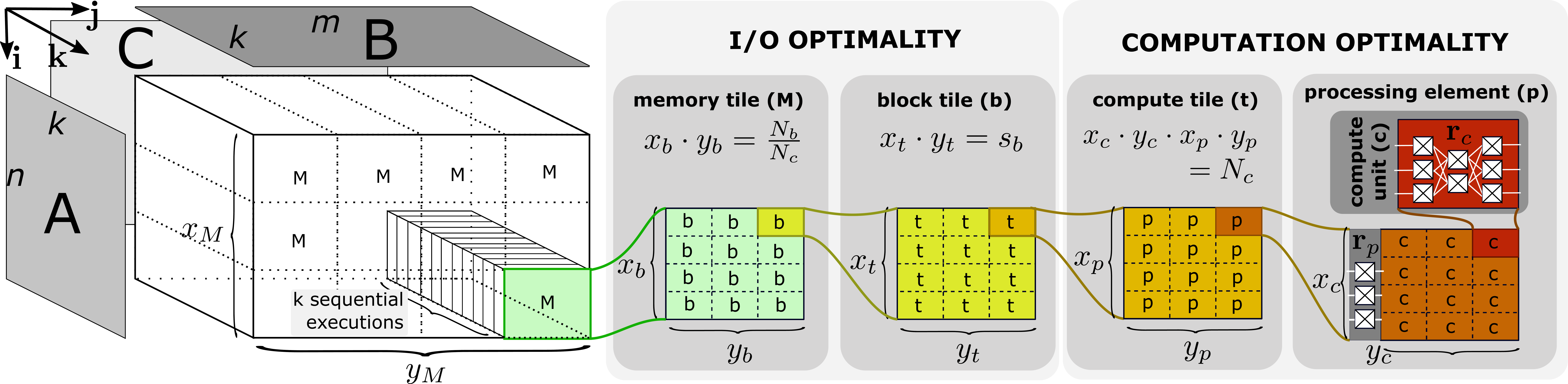}
  \caption{Decomposition of MMM achieving both performance and I/O optimality in
  terms of hardware resources.}
	\label{fig:memory_tiling_scheme}
\end{figure*}

\vspace{-0.75em}
\section{Introduction}
\vspace{-0.25em}
\label{sec:introduction}

The increasing technological gap between computation and memory
speeds~\cite{hitting_the_memory_wall} pushes both academia~\cite{25d, mmmSched,
communicationOptMMM, externalMem} and industry~\cite{intel2007intel,
anderson1999lapack} to develop algorithms and techniques to minimize data
movement. The first works proving I/O lower bounds for specific algorithms,
e.g., Matrix Matrix Multiplication (MMM), date back to the
80s~\cite{red_blue_pebble_game}.  The results were later extended to parallel
and distributed machines~\cite{toledo}. Since then, many I/O minimizing
algorithms were developed for linear algebra~\cite{25d, Cholesky, QR}, neural
networks~\cite{comm_DNN}, and general programs that access
arrays~\cite{general_arrays}.
Minimizing I/O impacts not only performance, but also reduces bandwidth usage in
a shared system.  MMM is typically used as a component of larger
applications~\cite{DNN-MMM, joost}, where it co-exists with other algorithms,
e.g., memory bound linear algebra operations such as matrix-vector/vector-vector
operations, which benefit from a larger share of the bandwidth, but do not
require large amounts of compute resources.

FPGAs are an excellent platform for accurately modeling performance and I/O to
guide algorithm implementations. In contrast to software implementations, the
replacement of cache with explicit on-chip memory, and isolation of the
instantiated architecture, yields fully deterministic behavior in the circuit:
accessing memory, both on-chip and off-chip, is always done explicitly, rather
than by a cache replacement scheme fixed by the hardware. 
The models established so far, however, pose a challenge for their applicability
on FPGAs. They often rely on abstracting away many hardware details, assuming
several idealized processing units with local memory and all-to-all
communication~\cite{25d, externalMem, red_blue_pebble_game, toledo}.  Those
assumptions do not hold for FPGAs, where the physical area size of
custom-designed processing elements (PEs) and their layout are among most
important concerns in designing efficient FPGA
implementations~\cite{rapidwright}.  Therefore, performance modeling for
reconfigurable architectures requires taking constraints like logic resources,
fan-out, routing, and on-chip memory characteristics into account. 

With an ever-increasing diversity in available hardware platforms, and as
low-precision arithmetic and exotic data types are becoming key in modern
DNN~\cite{dnn_survey} and linear solver~\cite{MPLU} applications,
extensibility and flexibility of hardware architectures will be crucial to stay
competitive.  Existing high-performance FPGA implementations~\cite{gemm_intel,
jovanovic2012fpga} are implemented in hardware description languages (HDLs),
which drastically constrains their maintenance, reuse, generalizability, and
portability.  Furthermore, the source code is not disclosed, such that
third-party users cannot benefit from the kernel or build on the architecture.

In this work, we address all the above issues. We present a high-performance,
communication avoiding MMM algorithm, which is based on both computational
complexity theory~\cite{red_blue_pebble_game}
(Section~\ref{sec:io_optimal_schedule}), and on detailed knowledge of
FPGA-specific features (Section~\ref{sec:implementation}). Our architecture is
implemented in pure C++ with a small and readable code base, and to the best of
our knowledge, is the first open source, high-performance MMM FPGA code. We do
not assume the target hardware, and allow easy configuration of platform, degree
of parallelism, buffering, data types, and matrix sizes, allowing kernels to be
specialized to the desired scenario. 
\noindent The contributions of this paper are:
\begin{itemize}
  \item We model a decomposition for matrix multiplication that simultaneously
  targets maximum performance and minimum off-chip data movement, in terms of
  hardware constants.
  \item We design a mapping that allows the proposed scheme to be implemented in
  hardware, using the model parameters to lay out the architecture. 
  \item We provide a plug-and-play, open source implementation of the hardware
  architecture in pure HLS C++, enabling portability across FPGA and
  demonstrating low code complexity. 
  \item We include benchmarks for a wide range of floating point and integer
  types, and show the effect of adjusting parallelism and buffer space in the
  design, demonstrating the design's flexibility and scalability.
\end{itemize}

\section{Optimization Goals}
\label{sec:optimization_goals}

In this section we introduce \emph{what} we optimize. In
Sections~\ref{sec:io_optimal_schedule},~\ref{sec:tiling},
and~\ref{sec:implementation} we describe \emph{how} this is achieved.
We consider optimizing the schedule of a \emph{classical} MMM algorithm, that
is, given a problem of finding $C$, where
$C = AB,
A \in \mathbb{R}^{m \times k},
B \in \mathbb{R}^{k \times n},
C \in \mathbb{R}^{m \times n}$, an algorithm performs $F = mnk$
multiplications and additions (pseudocode shown in \lstref{mmm}). We therefore 
exclude
Strassen-like routines~\cite{Strassen} from our analysis, as the classical
algorithms often perform better on practical problems and
hardware~\cite{strassenVsClassic}. 
We require that the optimal schedule: (1) achieves \emph{highest performance}
(takes least time-to-solution time), while (2) performing the \emph{least number
of I/O operations}, by (3) \emph{making most efficient use of resources}.

\begin{figure}[h]
  \centering
  \begin{minipage}{.5\textwidth}
  \centering
  \begin{lstlisting}
for (i = 0; i < M; i++) 
    for (j = 0; j < N; j++) 
        for (k = 0; k < K; k++) 
            C[i, j] = C[i, j] + A[i, k]*B[k, j];
  \end{lstlisting}
  \vspace{-1em}
  \captionof{listing}{Classical MMM algorithm.}
  \label{lst:mmm}
  \end{minipage}
\end{figure}

\paragraph{Computation}
On general purpose CPUs, as well as on modern GPUs, optimizing for computation
is often a straightforward task. Exploiting techniques like vectorization or
data alignment~\cite{compilerVect} can mostly be offloaded to compilers. Thus,
most of the effort is spent on I/O minimization.  When targeting FPGAs,
designing an architecture that can optimally exploit available logic, even for
computationally simple algorithms such as matrix multiplication, requires
significant engineering effort. We must thus maintain a decomposition that is
efficiently implementable in hardware, while achieving the desired theoretical
properties.

\vspace{0.5em}

\paragraph{I/O}
Schedule optimization on a parallel machine determines both the domain
decomposition (which computations are assigned to which compute units), and
sequential execution (the order in which each compute unit executes its tasks).
The former impacts communication between compute units (a.k.a.  \emph{horizontal
I/O}), and the latter is responsible for the communication between a compute
unit and a main memory (a.k.a. \emph{a vertical I/O}). Both aspects of the
parallel schedule are constrained by available resources and their
interdependencies (e.g., NUMA domains or limited fan-out on FPGAs).

\vspace{0.5em}

\paragraph{Resources}
\label{sec:fpga_routing}
When targeting a high utilization design on FPGA, it is critical to maintain
characteristics that aid the routing process.  Routing reacts poorly to large
fan-in or fan-out, which typically occurs when these are dependent on the degree
of parallelism: that is, if $N$ determines the degree of parallelism in the
program, $1$-to-$N$ and $N$-to-$1$ connections in the architecture should be
avoided.  This is true both on the granularity of individual logic units, and on
the granularity of coarse-grained modules instantiated by the programmer.  To
accommodate this, we can regulate the size of PEs, and favor PE topologies that
are easily mapped to a plane, such as grids or chains.
Furthermore, mapping of a hardware architecture to the chip logic and
interconnect (placement and routing) may reduce the clock frequency due to long
routing paths. Due to the intractable size of the configuration space, this
cannot be efficiently modeled and requires empirical evaluation of designs. The
routing challenges are exasperated in FPGA chips that consist of multiple
``chiplets'', such as the Xilinx UltraScale+ VU9P chip used in this paper, which
hosts three ``super-logical regions'' (SLRs). Crossing the chiplets consumes
highly limited routing resources and carries a higher timing penalty. Limiting
these crossings is thus key to scaling up resource utilization.

\begin{table}
	\setlength{\tabcolsep}{2pt}
	\renewcommand{\arraystretch}{0.7}
	\centering
	\small
	\sf
	\begin{tabular}{@{}l|ll@{}}
			\toprule
			\multirow{4}{*}{\begin{turn}{90}\textbf{MMM}\end{turn}}
			& $\vec{i}$, $\vec{j}$, $\vec{k}$ & Unit vectors spanning 3D 
			iteration
			space. \\
			& $m$, $n$, $k$& Matrix sizes in $\vec{i}$, $\vec{j}$, and $\vec{k}$
			dimensions, respectively. \\
			& $\matr{A}$, $\matr{B}$ & Input matrices $\matr{A} \in 
			\mathbb{R}^{m \times k}$
			and $\matr{B} \in \mathbb{R}^{k \times n}$. \\
			& $\matr{C} = \matr{A}\matr{B}$& Output matrix $\matr{C} \in 
			\mathbb{R}^{m \times
				n}$. \\
			\midrule
			\multirow{4}{*}{\begin{turn}{90}\textbf{naming}\end{turn}}
			& $\alpha_\beta$ & Parameter naming convention. $\alpha$ is some 
			quantity \\
			& & (i.e., size or number of objects), $\beta$ is an object that \\ 
			& &	$\alpha$ refers to. \\
			& $\alpha_{\beta, \text{max}}$ & Hardware limit on a parameter
			$\alpha_\beta$. \\
			& $\alpha_{\text{tot}} = \prod_{\beta} \alpha_\beta$ & The product 
			of all
			tile sizes. \\
			\midrule
			\multirow{5}{*}{\begin{turn}{90}$\bm{\alpha}$\end{turn}}
			& $N$&Total number of objects. \\
			& $x, y$ &Number of objects in $\mathbf{i}$ (or $\mathbf{j}$) 
			dimension in the \\
			& & enveloping tile. \\
			& $s$ & Intrinsic size of an object. \\
			& $w$ & Bit width of object (e.g., of port or data type). \\
			& $\vec{r}$ & Vector of logic resources. \\
			\midrule
			\multirow{4}{*}{\begin{turn}{90}
					$\bm{\beta}$
			\end{turn}} 
			& $c$ & Compute units in a processing element. \\
			& $p$ & Processing elements in a compute tile.  \\
			& $t$ & Compute tiles in a block tile. \\
			& $b$ & Block tiles in a memory tile. \\
			\midrule
			\multirow{6}{*}{\begin{turn}{90}
					\textbf{optimization}
			\end{turn}} 
			& $S = N_b \cdot s_b$ & Total size of available on-chip memory. \\ 
			& $N_c \le N_{c, \text{max}}$ & Total number of compute units. \\
			& $Q$ & Total number of off-chip memory transfers. \\
			& $F = n \cdot m \cdot k$ & Total number of multiply-addition 
			operations\\
			& &  required to perform MMM.\\	
			& $f \le f_{\text{max}}$ & Achieved and maximum design frequency. \\
			& $T = \frac{F}{f \cdot N_c}$ & Design total execution time. \\
			
			\bottomrule
	\end{tabular}
  \vspace{1em}
	\caption{The most important symbols used in this paper.}
	\label{tab:symbols}
  \vspace{-4em}
\end{table}

\vspace{0.5em}

\label{sec:compute_resources}
Throughout this paper, we use the two-level notation for naming parameters
(Table~\ref{tab:symbols}). Most of the parameter names are in the form of
$\alpha_\beta$, where $\alpha$ refers to some quantity, such as the total number
of objects, and $\beta$ determines what is the object of interest. E.g., $N_c,
N_b$, $s_b$ are: total number of ($N$) compute units ($c$), memory blocks ($b$),
and a size of each memory block ($s$), respectively.

The target hardware contains $d$ types of different logic resources. This
typically consists of general purpose logic, such as \emph{lookup tables}
(LUTs), and more specialized arithmetic units, such as \emph{digital signal
processing units} (DSPs). We represent a quantity of these resources as a
vector $\vec{r}_\text{max} = [r_{1,\text{max}}, ..., r_{d,\text{max}}]$. As a
basic logical entity, we consider a ``compute unit'', which is a basic circuit
able to perform a single multiply-addition operation in a single cycle. Each
unit is implemented on the target hardware using some combination of logic
resources $\vec{r}_c$. Depending on the numerical precision, different number
of computation resources $\vec{r}_c$ are needed to form a single compute unit,
so the maximum number of compute units that can be instantiated, $N_c$, may
vary.  The compute units are organized into $N_p$ \emph{processing elements}
(PEs), which encapsulate a closed logical task (e.g., a vector operation) of
$x_c \cdot y_c$ compute units. Each PE additionally consumes $\vec{r}_p$ logic
resources as orchestration overhead. This gives us the following constraint,
which enforces that the total amount of resources consumed by compute units and
their encompassing PEs should not exceed the total resources available: 
\begin{align}
  \begin{split}
  \forall_{1 \le i \le d}N_c r_{i,c} + N_p r_{i,p} & \le r_{i,\text{max}}\text{,} \\
  \text{or equivalently }\forall_{1 \le i \le d} N_p (r_{i,p} + r_{i,c} \cdot x_c y_c ) & \le 
r_{i,\text{max}} \\
  \end{split}
\end{align}
where $d$ is the dimensionality of the resource vector (illustrated in the
top-right side of \figref{memory_tiling_scheme}).

\section{Optimization Models}
\subsection{Computation Model}
\label{sec:performance_model}

To optimize computational performance we minimize the total execution runtime,
which is a function of achieved parallelism (total number of compute units
$N_c$) and the design clock frequency $f$. The computational logic is organized
into $N_p$ PEs, and we assume that every PE holds $x_c \cdot y_c$ compute units
in dimensions $\mathbf{x}$ and $\mathbf{y}$ (see \tabref{symbols} for an
overview of all symbols used).  We model the factor $N_c$ directly in the
design, and rely on empirically fixing $f$, which is limited by the maximum size
of data buses between PEs (i.e., $x_c w_c \leq w_{p,\text{max}}$ and $y_c w_c
\leq w_{p,\text{max}}$, where $w_{p,\text{max}}$ depends on the architecture,
and typically takes values up to $\SI{512}{\bit}$).
Formally, we can write the computational optimization problem as follows:

\begin{align}
\begin{split}
\text{minimize } T = \frac{F}{f \cdot N_c} &= \frac{mnk}{f \cdot N_p \cdot x_c
y_c} \\
\text{subject to:} \\
\forall_{1 \le i \le d} N_p (r_{i,p} + r_{i,c} \cdot x_c \cdot y_c ) & \le 
r_{i,\text{max}} \\
x_c w_c & \le w_{p,\text{max}} \\
y_c w_c & \le w_{p,\text{max}} \\
f & \le f_{\text{max}}
\end{split}
\label{eq:compute_opt}
\end{align}
That is, the time to completion $T$ is minimized when $f{\cdot}N_c$ is
maximized, where the number of parallel compute units $N_c$ is constrained by
the available logic resources $\vec{r}_\text{max}$ of the design (this can be
the full hardware chip, or any desired subset resource budget). We respect
routing constraints by observing a maximum bus width $w_{p,\text{max}}$, and
must stay within the target frequency $f_\text{max}$.

\subsection{I/O Model}
\label{sec:io_optimal_schedule}

\subsubsection{State-of-the-art of Modeling I/O for MMM}\leavevmode

\noindent Minimizing the I/O cost is essential for achieving high performance on
modern architectures, even for traditionally compute-bound kernels like
MMM~\cite{cosma}.  In this section, we sketch a theoretical background from
previous works which lays foundation for our FPGA model.
%
Following the state-of-the-art I/O
models~\cite{red_blue_pebble_game, IronyMMM, 25d} we assume that a parallel
machine consists of $p$ processors, each equipped with a fast private memory of
size $S$ words. To perform an arithmetic operation, processor $p_i$ is required
to have all operands in its fast memory.
The principal idea behind the I/O optimal schedule is to maximize the
\emph{computational intensity}, i.e., the number of arithmetic operations
performed per one I/O operation. This naturally expresses the notion of data
reuse, which reduces both vertical (through memory hierarchy) and horizontal
(between compute units) I/O (Section~\ref{sec:optimization_goals}).

\paragraph{Algorithm as a Graph}
We represent an entire execution of an algorithm as a \emph{computation directed
acyclic graph} (CDAG)~\cite{registerpebblecolor, red_blue_pebble_game, cosma}
$G=(V,E)$, where every vertex $v \in V$ corresponds to some unique value during
the execution, and edges $e \in E$ represent data dependencies between them.
Vertices without incoming edges are \emph{inputs}, and the ones without outgoing
edges are \emph{outputs}. The remaining vertices are intermediate results. In
the context of MMM, matrices $A$ and $B$ form $m \times k$ and $k \times n$
input vertices, respectively, and partial sums of $C$ form $m n k$ intermediate
vertices, with inputs both from corresponding vertices of $A$ and $B$, and
previous partial sums of $C$. The output vertices are formed by the $m \times n$
vertices of $C$ which represent the last of $k$ partial sums. The MMM CDAG is
shown in~\figref{mmm_cdag}a.

\paragraph{I/O as Graph Pebbling}
Hong and Kung~\cite{red_blue_pebble_game} introduced the red-blue pebble game
abstraction to model the I/O cost of a sequential schedule on a CDAG. We refer a
reader to the original paper for the formal definition and details of this game:
here we just draw its simplistic sketch. The key idea is to play a pebbling game
with a limited number of red pebbles (corresponding to the small-but-fast
memory) and an unlimited number of blue pebbles (large, slow memory). The rules
are that one can put a red pebble on a vertex only if all its direct
predecessors also have red pebbles (which represent computing a value, while all
operands are in the fast memory). Placing a blue pebble on a red one corresponds
to a store operation, and putting a red pebble on a blue corresponds to a load.
Initially, only input vertices have blue pebbles on them. The goal is to put
blue pebbles on the output vertices. In this model, the I/O optimal schedule is
a sequence of pebbling moves which minimizes the load and store operations.

\paragraph{Schedule as a Graph Partition}
In our work, we extend the methodology and results of COSMA~\cite{cosma}, which
is based on the red-blue pebble game.  We partition the CDAG $G = (V,E)$ into
$h$ disjoint subsets (a.k.a. \emph{subcomputations}) $V_i, i = 1 \dots h, V_i
\subset V$, such that each $V_i$ has a constant number $X$ of input and output
vertices (which form the \emph{Dominator set} and \emph{Minimum set},
respectively). The collection of all $\{V_i\}, \bigcup V_i = V$ is called an
$X$-partition of $G$.  The authors show that an I/O optimal scheme can be
derived from finding an $X$-partition $\{V_i\}$ of a smallest cardinality, for
some value of $X$. We will use this result to build our model.

\paragraph{Optimal Graph Partition as Maximizing Computational Intensity}
\label{sec:io_model}
In COSMA~\cite{cosma}, it is shown that the I/O optimal MMM schedule maximizes
the \emph{computational intensity} of each subcomputation $V_i$, that is, the
number  of arithmetic operations per I/O operation. Formally:

\begin{align}
\begin{split}
\text{maximize } & \frac{|V_i|}{|\text{Dom}(V_i)| - |V_{R,i}| + |W_{B,i}|} \\
\text{subject to: } & |V_{R,i}| \le S\text{,}
\label{eq:memory_opt}
\end{split}
\end{align}

\noindent where $|V_i|$ is a number of values updated in subcomputation $V_i$,
$|\text{Dom}(V_i)|$ is the number of inputs of $V_i$ (the \emph{Dominator set}),
$|V_{R,i}|$ is the number of inputs that are already in the on-chip memory (data
reuse), and $|W_{B,i}|$ is the number of partial results that have to be stored
back to off-chip memory.  Therefore, we aim to maximize utilization of all
available on-chip memory, to increase the reuse of data $|V_{R,i}|$ that is
already loaded. The sets $V_i$, $Dom(V_i)$, and $V_{R,i}$ are depicted
in~\figref{mmm_cdag}b.

\subsubsection{Extending to the FPGA Scenario}\leavevmode

\paragraph{I/O Optimal Schedule vs. FPGA Constraints}
State-of-the-art I/O models~\cite{red_blue_pebble_game, IronyMMM, 25d} assume
that a parallel machine consists of $p$ processors, each equipped with a small
private memory of constant size $S$ words (two-level memory model). Under these
assumptions, COSMA establishes that the I/O optimal MMM schedule is composed of
$h = mnk/S$ subcomputations, each performing $S$ multiply-addition operations
while loading $2\sqrt{S}$ elements from matrices $A$ and $B$.  However, in the
FPGA setting these assumptions do not hold, as the number of compute units $N_c$
is a variable depending on both hardware resources (which is constant), and on
the processing element design. Furthermore, the mapping between processing
elements and available BRAM blocks is also constrained by ports and limited
fan-out.
We impose the additional requirement that compute and memory resources must be
equally distributed among PEs, posing additional restrictions on a number of
available resources and their distribution for each subcomputation $V_i$ to
secure maximum arithmetic throughput and routing feasibility:

\begin{enumerate}
  \item The number of parallel compute units $N_c$ is maximized.
  \item The work is load balanced, such that each compute unit performs the same
  number of computations.
  \item Each memory block is routed to only one compute unit (i.e., they are not
  shared between compute units).
  \item Each processing element $p$ performs the same logical task, and consumes
  the same amount of computational and memory block resources.
\end{enumerate}

\paragraph{Memory resources}
\label{sec:memory_resources}
To model the memory resources of the FPGA, we consider the bit-length of $w_c$,
depending on the target precision. The machine contains $N_b$ on-chip memory
blocks, each capable of holding $s_b$ words of the target data type, yielding a
maximum of
\begin{align*}
  S = N_b \cdot s_b
\end{align*}
words that can be held in on-chip memory. $s_b$ takes different values depending
on $w_c$ (e.g., $\SI{16}{\bit}$ for half precision floating point, or a
$\SI{64}{\bit}$ long unsigned integer). Each memory block supports one read and
one write of up to $w_b$ bits in a single cycle in a pipelined fashion.

\paragraph{FPGA-constrained I/O Minimization}
We denote each $V_i$ as a 
\emph{memory
tile} $M$, as its size in $\vec{i}$ and $\vec{j}$ dimensions determines the 
memory reuse.
To support a hierarchical hardware design, each $M$ is further decomposed into
several levels of tiling.  This decomposition encapsulates hardware features of
the chip, and imposes several restrictions on the final shape of $M$. The tiling
scheme is illustrated in \figref{memory_tiling_scheme}.
We will cover the purpose and definition of each layer in the hierarchy shortly
in \secref{resource_model}, but for now use that the dimensions of the full
memory tile $M$ are:
\begin{align}
  \begin{split}
    x_{\text{tot}} &= x_c \cdot x_p \cdot x_t \cdot x_b \\
    y_{\text{tot}} &= y_c \cdot y_p \cdot y_t \cdot y_b\text{, }
  \end{split}
\end{align}
and we set $|V_i| = x_{\text{tot}}y_{\text{tot}}$.  Following the result
from~\cite{cosma}, a schedule that minimizes the number I/O operations, loads
$x_{\text{tot}}$ elements of one column of matrix $\matr{A}$, $y_{\text{tot}}$
elements of one row of matrix $\matr{B}$ and reuses
$x_{\text{tot}}y_{\text{tot}}$ previous partial results of $\matr{C}$, thus
computing an outer product of the loaded row and column. We now rewrite
\equref{memory_opt} as:

\begin{align}
  \begin{split}
    \text{maximize } &\frac{x_{\text{tot}}y_{\text{tot}}}{x_{\text{tot}}
    + y_{\text{tot}}} \\
    \text{subject to: } & x_{\text{tot}} + y_{\text{tot}} \le S \\
    & x_{\text{tot}}y_{\text{tot}}  \le S\text{, }
  \end{split}
\label{eq:memory_opt_fpga}
\end{align}
and the total number of I/O operations as:
\begin{align}
  Q = mn\left(1 + k\left(\frac{1}{x_{\text{tot}} } +
  \frac{1}{y_{\text{tot}} }\right)\right)\text{.}
  \label{eq:comm_vol}
\end{align}
This expression is minimized when:
\begin{align}
  x_{\text{tot}} = y_{\text{tot}} = \sqrt{S}
  \label{eq:square_tile}
\end{align}
That is, a memory tile is a square of size $S$.  \equref{comm_vol} therefore
gives us a theoretical lower bound on $Q \le 2mnk / \sqrt{S}$, assuming that all
available memory can be used effectively. However, the assumptions stated in
\secref{io_optimal_schedule} constrain the perfect distribution of hardware
resources, which we model in~\secref{tiling}.

\begin{figure}[h]
  \begin{minipage}{.45\textwidth}
    \begin{lstlisting}[breaklines=false]
// Memory tiles $m$
for ($i_m = 1$; $i_m \le n$; $i_m = i_m + x_{\text{tot}}$) 
  for ($j_m = 1$; $j_m \le m$; $j_m = j_m + y_{\text{tot}}$)  
    for ($k = 1$; $k \le k$; $k = k + 1$) // Full dimension $k$
// [Sequential] Block tiles $b$ in memory tile
      for ($i_b = i_m$; $i_b \le i_m + x_\text{tot}$; $i_m = i_m + x_t x_c x_p$)
        for ($j_b = j_m$; $j_b \le j_m + y_\text{tot}$; $j_m = j_m + y_t y_p y_c$)
// [Sequential] Compute tiles $t$ in block tile 
          for ($i_t = i_b$; $i_t \le i_b + x_t x_p x_c$; $i_b = i_b + x_c x_p$)
            for ($j_t = y_b$; $j_t \le j_b + y_t y_p y_c$; $j_t = j_t + y_c y_p$)
// [Parallel] Processing elements $p$ in compute tile 
              forall ($i_p = i_t$; $i_p \le i_t + x_p x_c$; $i_t = i_t + x_c$)
                forall ($j_p = j_t$; $j_p \le j_t + y_p y_c$; $j_p = j_p + y_c$)
// [Parallel] Compute units $c$ in processing element 
                  forall ($i_c = i_p$; $i_c \le i_p + x_c$; $i_c = i_c + 1$)
                    forall ($j_c = j_p$; $j_c \le j_p + y_c$; $j_c = j_c {+} 1$)
                      $\matr{C}(i_c, j_c) = \matr{C}(i_c, j_c) + \matr{A}(i_c,
                      k) \cdot \matr{B}(k, j_c)$
    \end{lstlisting}
    \vspace{-1em}
    \captionof{listing}{Pseudocode of the tiled MMM algorithm.}
    \label{lst:pseudocode}
  \end{minipage}
\end{figure}

\subsection{Resource Model}
\label{sec:tiling}
\label{sec:resource_model}

Based on the I/O model and the FPGA constraints, we create a logical hierarchy
which encapsulates various hardware resources, which will guide the
implementation to maximize I/O and performance.
We assume a chip contains $\vec{r}_\text{max} =
\{r_{1,\text{max}}, \dots, r_{t,\text{max}}\}$ different hardware resources (see
\secref{compute_resources}). The dimensionality and length of a vector depends
on the target hardware -- e.g., Intel Arria~10 and Stratix~10 devices expose
native floating point DSPs, each implementing a single operation, whereas a
Xilinx UltraScale+ device requires a combination of logic resources.
We model fast memory resources separately as memory blocks (e.g., M20K
blocks on Intel Stratix~10, or Xilinx BRAM units). We consider a chip
contains $N_b$ memory blocks, where each unit can store $s_b$ elements of the
target data type and has a read/write port width of $w_b$ bits.
The scheme is organized as follows (shown in \figref{memory_tiling_scheme}):
\begin{enumerate}
  \item A compute unit $c$ consumes $\vec{r}_c$ hardware resources, and can
  throughput a single multiplication and addition per cycle.  Their maximal
  number 
  $$N_{c,\text{max}} \le \min_{1 \le i \le t}
  \Big(\frac{r_{i,\text{max}}}{r_{i,c}}\Big)$$ 
  for a given numerical precision is a hardware constant, given by the available
  resources $\vec{r}_{\text{max}}$.
  \item A processing element $p$ encapsulates $x_c \cdot y_c$ compute units.
  Each processing element requires additional $\vec{r}_p$ resources for overhead
  logic.
  \item A compute tile $t$ encapsulates $x_p \cdot y_p$ processing elements.
  One compute tile contains all available compute units \\$x_c \cdot y_c \cdot x_p
  \cdot y_p = N_c$.
  \item A block tile $b$ encapsulates $x_t \cdot y_t = s_b$ compute tiles,
  filling the entire internal capacity $s_b$ of currently allocated memory
  blocks.
  \item A memory tile $M$ encapsulates $x_b \cdot y_b =
  \floor*{\frac{N_b}{N_{b,\text{min}}}}$ block tiles (discussed below), using
  all available $N_b$ memory blocks.
\end{enumerate}
Pseudocode showing the \emph{iteration space} of this decomposition is shown in
\lstref{pseudocode}, consisting of 11 nested loops. Each loop is either a
\emph{sequential} \texttt{for}-loop, meaning that no iterations will overlap,
and will thus correspond to \emph{pipelined} loops in the HLS code; or a
\emph{parallel} \texttt{forall}-loop, meaning that every iteration is executed
every cycle, corresponding to \emph{unrolled} loops in the HLS code. We require
that the sequential loops are coalesced into a single pipeline, such that no
overhead is paid at iterations of the outer loops. \\

\subsection{Parallelism and Memory Resources}

The available degree of parallelism, counted as a number of simultaneous
computations of line~17 in Listing~\ref{lst:pseudocode}, is determined by the
number of compute units $N_c$. Every one of these compute units must read and
write an element of $\matr{C}$ from fast memory \emph{every cycle}. This implies
a minimum number of \emph{parallel} fast memory accesses that must be supported
in the architecture. 
Memory blocks expose a limited access width $w_b$ (measured in bits), which
constrains how much data can be read from/written to them in a single cycle. We
can thus infer a minimum number of memory blocks necessary to serve all compute
units in parallel, given by:
\begin{align}
  N_{b,\text{min}} = x_p y_p \cdot \ceil*{\frac{w_c \cdot x_c y_c}{w_b}}\text{,}
  \label{eq:port_width}
\end{align}
where $w_c$ is the width of the data type in bits, and $x_c y_c$ denotes the
granularity of a processing element. Because all $x_c y_c$ accesses within a
processing element happen in parallel, accesses to fast memory can be coalesced
into long words of size $w_c \cdot x_c y_c$ bits. For cases where $w_b$ is not a
multiple of $w_c$, the ceiling in \equref{port_width} may be significant for the
resulting $N_{b,\text{min}}$. When instantiating fast memory to implement the
tiling strategy, \equref{port_width} defines the minimum ``step size'' we can
take when increasing the tile sizes. 

\begin{figure}
  \centering
  \includegraphics[width=\columnwidth]{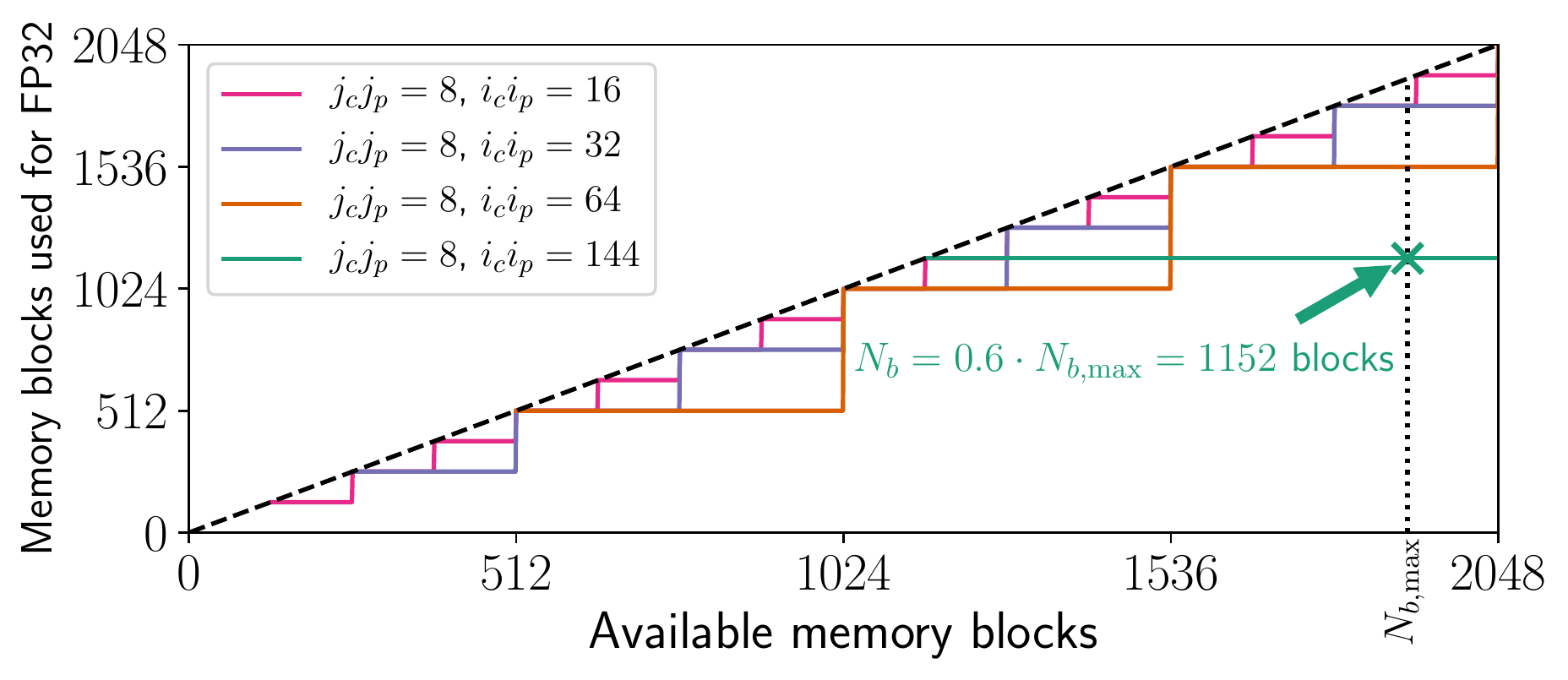}
  \vspace{-2em}
  \caption{Utilization of memory blocks with memory tile size. For $i_c j_c=8$
  and $i_p j_p=144$, we can utilize $60.4\% \cdot N_{b,\text{max}}$.}
  \label{fig:ram_usage}
\end{figure}

Within a full memory tile, each updated value $C[i,j]$ is reused after all
$x_{\text{tot}} \cdot y_{\text{tot}}$ elements in a single memory tile are
evaluated, and computation proceeds to the next iteration of the $k$-loop
(line~4 in Listing~\ref{lst:pseudocode}).  Given the intrinsic size of each
memory block $s_b$, we can thus perform $s_b$ iterations of the compute tile
before a single batch of $N_{b,\text{min}}$ allocated memory blocks has been
filled up. If the total number of memory blocks $N_{b,\text{max}} \ge 2
N_{b,\text{min}}$, i.e., the number of blocks required to support the parallel
access requirements is less than the total number of blocks available, we can
perform additional $\floor*{\frac{N_b}{N_{b,\text{min}}}}$ iterations of the
block tile, using all available memory blocks (up to the additive factor of $N_b
\mod N_{b,\text{min}}$).  However, for large available parallelism $N_c$, this
additive factor may play a significant role, resulting in a part of available
on-chip memory not being used.  This effect is depicted in \figref{ram_usage}
for different values of $N_c$ for the case of single precision floating point
(FP32) in Xilinx BRAM blocks, where $s_b = 1024$ and $w_b = \SI{36}{\bit}$. The
total number of memory blocks that can be efficiently used, without sacrificing
the compute performance and load balancing constraints, is then:
\begin{equation}
N_b = \floor*{\frac{N_{b,max}}{N_{b,\text{min}}}} N_{b,\text{min}}\text{.}
\label{eq:max_memory}
\end{equation}
In the worst case, this implies that only $N_{b,\text{max}}/2 + 1$ memory blocks
are used. In the best case, $N_{b,\text{max}}$ is a multiple of
$N_{b,\text{min}}$, and all memory block resources can be utilized.  When $N_c >
N_{b,\text{max}} / 2$, the memory tile collapses to a single block tile, and the
total memory block usage is equal to \equref{port_width}. 

\section{Hardware Mapping}
\label{sec:implementation}

With the goals for compute performance and I/O optimality set by the model, we
now describe a mapping to a concrete hardware implementation. 

\subsection{Layout of Processing Elements}

For \equref{compute_opt} to hold, all $N_p$ PEs must run at full 
throughout execution of the kernel, computing distinct contributions to the
output tile.
In terms of the proposed tiling scheme, we must evaluate a full compute tile $t$
(second layer in \figref{memory_tiling_scheme}) every cycle, which consists of
$x_p \cdot y_p$ PE tiles (first layer in \figref{memory_tiling_scheme}), each
performing $x_c \cdot y_c$ calculations in parallel, contributing a total of
$N_c$ multiplications and additions towards the outer product currently being
computed.
Assuming that $N_p$ elements of $\matr{A}$ and a full row of $y_\text{tot}$
elements of $\matr{B}$ have been prefetched, we must -- for each of the $x_p$
rows of the first layer in \figref{memory_tiling_scheme} -- propagate $x_c$
values to all $y_p$ horizontal PEs, and equivalently for columns of $\matr{B}$.
If this was broadcasted directly, it would lead to a total fan-out of $x_p \cdot
y_p$ for both inputs. 

\begin{figure}
  \centering
  \includegraphics[width=.7\columnwidth]{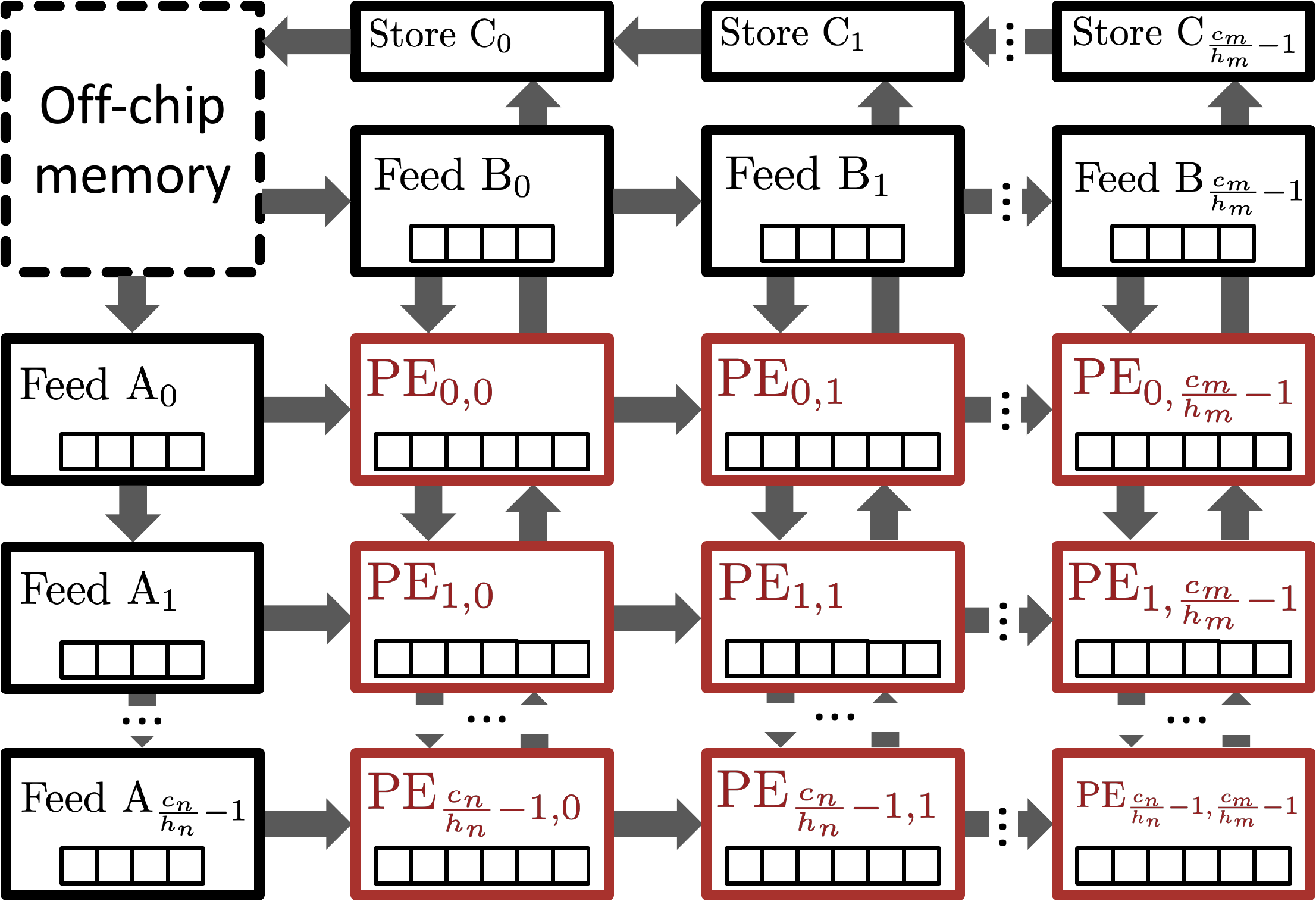}
  \caption{Compute arranged in a 2D grid.}
  \label{fig:systolic_2d}
\end{figure}

Rather than broadcasting, we can exploit the regular grid structure, letting
each column forward values of $\matr{A}$, and each row forward values of
$\matr{B}$, in a pipelined fashion. Such an architecture is sometimes referred
to as a \emph{systolic array}, and is illustrated in \figref{systolic_2d}. In
this setup, each processing element has three inputs and three outputs (for
$\matr{A}$, $\matr{B}$, and $\matr{C}$), and dedicated \texttt{Feed A} and
\texttt{Feed B} modules send prefetched contributions to the outer product at
the left and top edges of the grid, while \texttt{Store C} consumes the output
values of $\matr{C}$ written back by the PEs.  The number of inter-module
connections for this design is $3 x_p y_p$, but more importantly, the fan-out of
all modules is now constant, with $6$ data buses per PE.
Each PE is responsible for fully evaluating $x_\text{tot} y_\text{tot} / N_p$
elements of the output tile of $\matr{C}$. The elements of each PE tile in
\figref{memory_tiling_scheme} are stored contiguously (the first layer), but all
subsequent layers are not -- only the compute tile as a whole in contiguous in
$\matr{C}$. Final results must thus be written back in an interleaved manner to
achieve contiguous writes back to $\matr{C}$.

\paragraph{Collapsing to a 1D array.}
Although the 2D array of PEs is intuitive for performing matrix multiplication,
it requires a grid-like structure to be routed on the chip. While this solves
the issue of individual fan-out -- and may indeed be sufficient for monolithic
devices with all logic arranged in a rectangular structure -- we wish to map
efficiently onto general interconnects, including non-uniform and hierarchical
structures, as well as multiple-chiplet FPGAs (or, potentially, multiple FPGAs).
To achieve this, we can optionally collapse the 2D array of PEs into a 1D array
by fixing $y_p = 1$, resulting in $N_p = x_p$ PEs connected in sequence.  Since
this results in a long, narrow compute tile, we additionally fix $x_c = 1$,
relying on $y_c$ to regulate the PE granularity. Forming narrow compute tiles is
possible without violating \equref{square_tile}, as long as $x_\text{tot}$ and
$y_\text{tot}$ are kept identical (or as similar as possible), which we can
achieve by regulating the outer block and tiling layers (the memory and block
tile layers in \figref{memory_tiling_scheme}). 

\begin{figure}
  \centering
  \includegraphics[width=.7\columnwidth]{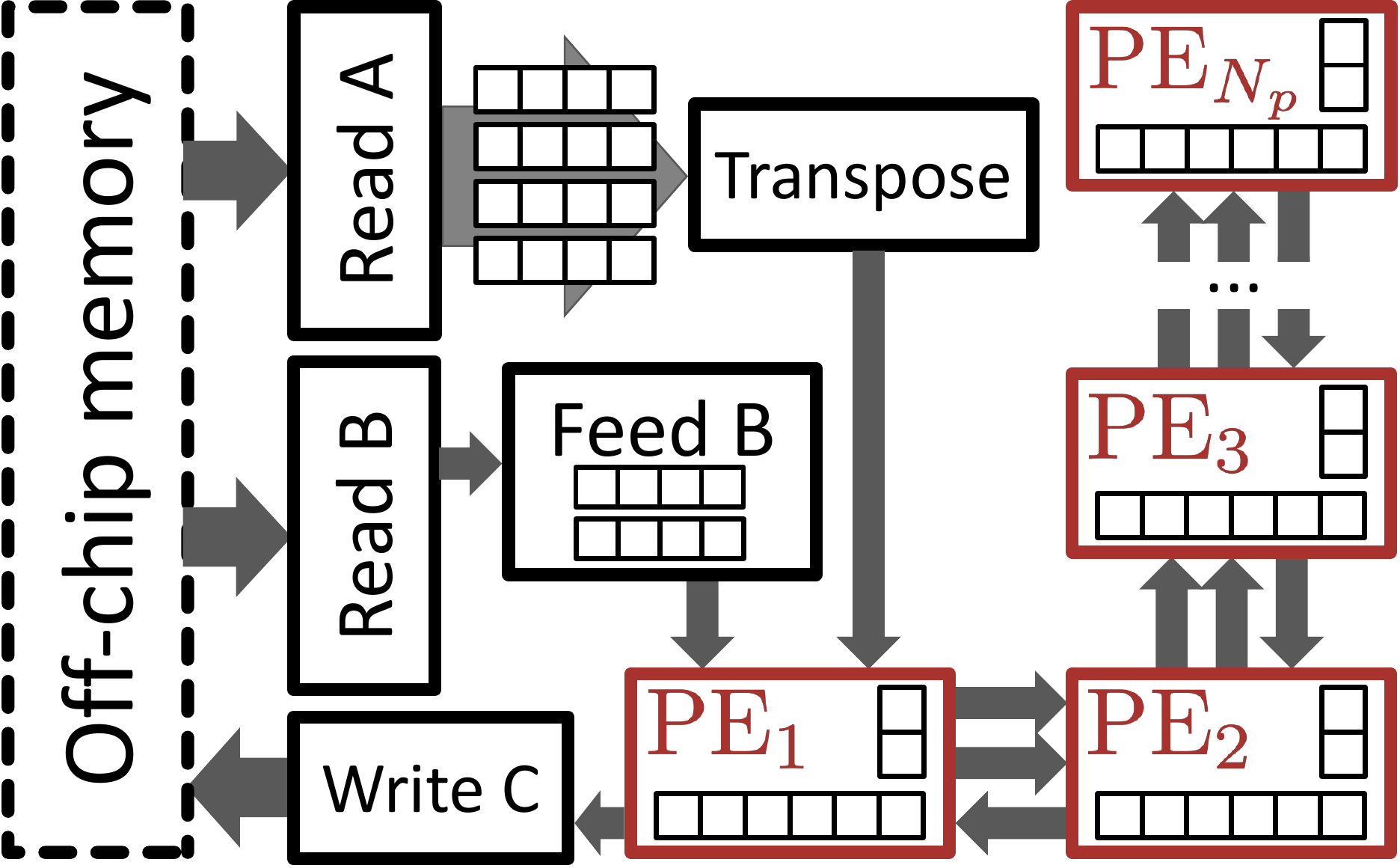}
  \caption{Module layout of final kernel architecture.}
  \label{fig:module_layout}
\end{figure}

\paragraph{Double buffering.}
Since each PE in the 1D array now computes one or more full rows of the compute
tile, we can buffer values of $\matr{A}$ in internal registers, rather than from
external modules. These can be propagated through the array from the first
element to the last, then kept in local registers and applied to values of
$\matr{B}$ that are streamed through the array from a buffer before the first
PE\@. Since the number of PEs in the final design is large, we overlap the
propagation of new values of $\matr{A}$ with the computation of the outer
product contribution using the previous values of $\matr{A}$, by using
\emph{double buffering}, requiring two registers per PE, i.e., $2 N_p$ total
registers across the design.

By absorbing the buffering of $\matr{A}$ into the PEs, we have reduced the
architecture to a simple chain of width $1$, reducing the total number of
inter-module connections for the compute to $3 N_p$, with 3 buses connecting
each PE transition. When crossing interconnects with long timing delays or
limited width, such as connections between chiplets, this means that only 3
buses must cross the gap, instead of a number proportional to the circumference
of the number of compute units within a single chiplet, as was the case for the
2D design. As a situational downside, this increases the number of pipeline
stages in the architecture when maximizing compute, which means that the number
of compute tiles must be larger than the number of PEs, i.e., $y_t x_t \geq N_p$.
This extra constraint is easily met when minimizing I/O, as the block tile size
is set to a multiple of $s_b$ (see \secref{tiling}), which in practice is higher
than the number of PEs, assuming that extreme cases like $x_c = y_c = 1$ are
avoided for large $N_c$.

\subsection{Handling Loop-carried Dependencies}

Floating point accumulation is often not a native operation on FPGAs, which can
introduce loop-carried dependencies on the accumulation variable. This issue is
circumvented with our decomposition. Each outer product consists of $x_p x_m
\cdot y_p y_m$ inner memory tiles. Because each tile reduces into a distinct
location in fast memory, collisions are separated by $x_p x_m \cdot y_p y_m$
cycles, and thus do not obstruct pipelining for practical memory tile sizes
(i.e., where $x_p x_m \cdot y_p y_m$ is bigger than the accumulation latency).

For data types such as integers or fixed point numbers, or architectures that
support (and benefit from) pipelined accumulation of floating point types, it is
possible to make $k$ the innermost loop, optionally tiling $n$ and $m$ further
to improve efficiency of reads from off-chip memory. The hardware architecture
for such a setup is largely the same as the architecture proposed here, but
changes the memory access pattern.

\subsection{Optimizing Column-wise Reads}

In the outer product formulation, the $\matr{A}$-matrix must be read in a
column-wise fashion. For memory stored as row-major, this results in slow and
wasteful reads from DDR memory (in a column-major setting, the same argument
applies, but for $\matr{B}$ instead).  For DDR4 memory, a minimum of 512 bits
must be transferred to make up for the I/O clock multiplier, and much longer
bursts are required to saturate DDR bandwidth in practice. To make up for this,
we can perform on-the-fly transposition of $\matr{A}$ as part of the hardware
design in an additional module, by reading wide vectors and pushing them to
separate FIFOs of depth ${\geq}x_b x_m$, which are popped in transposed order
when sent to the kernel (this module can be omitted in the implementation at
configuration time if $\matr{A}$ is pre-transposed, or an additional such module
is added if $\matr{B}$ is passed in transposed form).

\subsection{Writing Back Results}
\label{sec:writing_back_results}

Each final tile of $\matr{C}$ is stored across the chain of processing in a way
that requires interleaving of results from different PEs when writing it back to
memory. Values are propagated backwards through the PEs, and are written back to
memory at the head of the chain, ensuring that only the first PE must be close
to the memory modules accessing off-chip memory.
In previous work, double buffering is often employed for draining results, at
the significant cost of reducing the available fast memory from $S$ to $S/2$ in
\equref{comm_vol}, resulting in a reduction in the arithmetic intensity of
$\sqrt{2}$.
To achieve optimal fast memory usage, we can leave writing out results as a
sequential stage performed after computing each memory tile.  It takes $n m /
y_c$ cycles to write back values of $\matr{C}$ throughout kernel execution,
compared to $n m k / N_c$ cycles taken to perform the compute. When $k / N_c \gg
1$, i.e., the matrix is large compared to the degree of parallelism, this effect
of draining memory tiles becomes negligible.

\subsection{Final Module Layout}

With the constraints and considerations accumulated above, we fix the final
hardware architecture. The module layout is shown in \figref{module_layout}, and
consists of $4 + N_p$ modules.  The \texttt{Feed B} module buffers the outer
product row of $\matr{B}$, whereas $N_p$ values of $\matr{A}$ are kept in PE
registers. The vast majority of fast memory is spent in buffering the output
tile of $C$ (see \secref{io_optimal_schedule}), which is partitioned across the
PEs, with $\frac{x_\text{tot} \cdot y_\text{tot}}{N_p}$ elements stored in each.
The \texttt{Read A} and \texttt{Transpose} modules are connected with a series
of FIFOs, the number of which is determined by the desired memory efficiency in
reading $\matr{A}$ from DRAM\@. In our provided implementation, PEs are
connected in a 1D sequence, and can thus be routed across the FPGA in a
``snake-like'' fashion~\cite{rapidwright} to maximize resource utilization with
minimum routing constraints introduced by the module interconnect.

\begin{figure}
  \centering
  \includegraphics[width=.7\columnwidth]{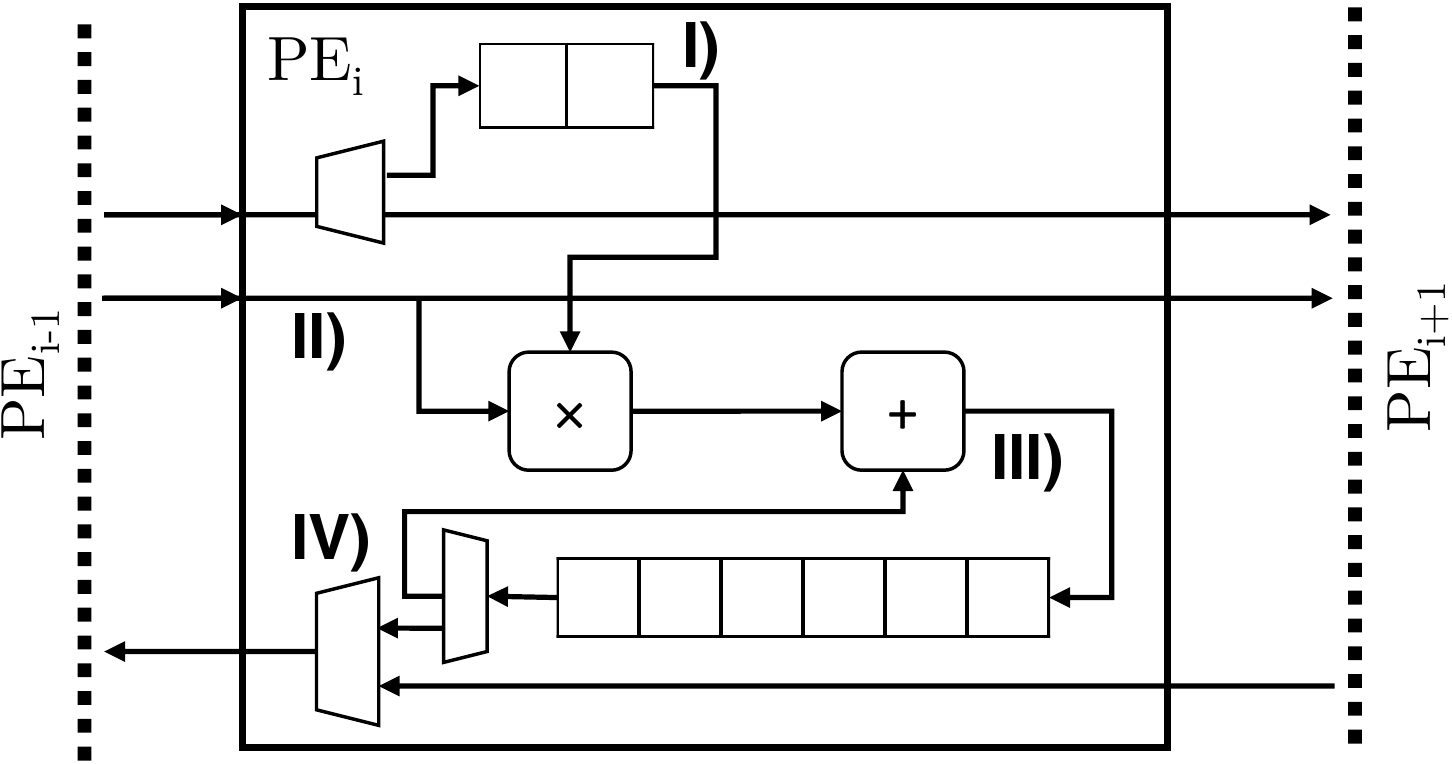}
  \caption{Architecture of a single PE.}
  \label{fig:pe}
\end{figure}

The PE architecture is shown in \figref{pe}. \textbf{I)} Each PE is responsible
for storing a single double-buffered value of $\matr{A}$.  Values are loaded
from memory and passed through the array, while the previous outer product is
being computed. \textbf{II)} Values of $\matr{B}$ are streamed through the chain
to be used at every PE. \textbf{III)} Every cycle accumulates into a different
address of the output $\matr{C}$ until it repeats after $x_t x_b \cdot y_t y_b$
cycles. \textbf{IV)} When the outer tile has been computed, it is sent back
through the PEs and written back at the memory interface.

\section{Evaluation}
\label{sec:evaluation}

\subsection{Parameter Selection}

Using the performance model and hardware mapping considerations, parameters for
kernel builds used to produce results are chosen in the following way, in order
to maximize performance and minimize I/O based on available compute and memory
resources, respectively:
\begin{enumerate}
  \item The PE granularity is fixed at $x_c = 1$, and $y_c$ is set as high as
  possible without impairing routing (determined empirically).  
  \item $f N_c$ is maximized by scaling up parallelism $N_c = N_p \cdot y_c$ (we
  fixed $x_c = 1$) when the benefit is not eliminated by reduction in frequency,
  according to \equref{compute_opt}.
  \item Memory tile sizes are maximized according to \equref{max_memory} to
  saturate on-chip memory resources. 
\end{enumerate}
For a given set of parameters, we build kernels in a \emph{fully automated
end-to-end fashion}, leveraging the abstractions provided by the high-level
toolflow. 

\subsection{Code Complexity}
\label{sec:code}

The MMM kernel architecture used to produce the result in this work is
implemented in Xilinx' Vivado HLS tool with hlslib~\cite{hlslib} extensions, and
as of writing this paper, consists of 624 and 178 SLOC of C++ for kernel and
header files, respectively. This is a generalized implementation, and includes
variations to support transposed/non-transposed input matrices, variable/fixed
matrix sizes, and different configurations of memory bus widths.  Additionally,
the operations performed by compute units can be specified, e.g., to compute the
distance product by replacing multiply and add with add and minimum.  The full
source code is available on github under an open source license (see footnote on
first page). 

\subsection{Experimental Setup}\sloppy

We evaluate our implementation on a Xilinx VCU1525 accelerator board, which
hosts an Virtex UltraScale+ XCVU9P FPGA. The board has four DDR4 DIMMs, but due
to the minimal amount of I/O required by our design, a single DIMM is sufficient
to saturate the kernel. The chip is partitioned into three chiplets, that have a
total of $\SI{1033608}{}$ LUTs, $\SI{2174048}{}$ \emph{flip-flops} (FFs),
$\SI{6834}{}$ DSPs, and $\SI{1906}{}$ BRAMs available to our kernels. This
corresponds to $87\%$, $92\%$, $99.9\%$, and $90\%$ of data sheet numbers,
respectively, where the remaining space is occupied by the provided shell.

Our kernels are written in Vivado~HLS targeting the
\texttt{xilinx:vcu1525:dynamic:5.1} platform of the SDAccel~2018.2 framework,
and \texttt{-O3} is used for compilation. We target $\SI{200}{\mega\hertz}$ in
Vivado~HLS and SDAccel, although this is often reduced by the tool in practice
due to congestion in the routed design for large designs, in particular paths
that cross between chiplets on the FPGA (see \secref{fpga_routing}).  Because of
the high resource utilization, each kernel build takes between $8$ and $24$
hours to finish successfully, or between $4$ and $24$ hours to fail placement or
routing.

On the Virtex UltraScale+ architecture, floating point operations are not
supported natively, and must be implemented using a combination of DSPs and
general purpose logic provided by the toolflow. The resource vector $\vec{r}$
thus has the dimensions LUTs, FFs, and DSPs. The Vivado~HLS toolflow allows
choosing from multiple floating point implementations, that provide different
trade-offs between LUT/FF and DSP usage. In general, we found that choosing
implementations of floating point addition that does not use DSPs yielded better
results, as DSPs replace little general purpose logic for this operation, and
are thus better spent on instantiating more multiplications.

Memory blocks are implemented in terms of BRAM, where each block has a maximum
port width of $\SI{36}{\bit}$ of simultaneous read and write access to
$\SI{18}{\kilo\bit}$ of storage. For wider data types, multiple BRAMs are
coalesced. Each BRAM can store $s_{b,\SI{36}{\bit}} = \SI{1024}{}$ elements in
$\SI{36}{\bit}$ configuration (e.g., FP32), $s_{b,\SI{18}{\bit}} = \SI{2048}{}$
elements in $\SI{18}{\bit}$ configuration (e.g., FP16), and $s_{b,\SI{72}{\bit}}
= 512$ elements in $\SI{72}{\bit}$ configuration (e.g., FP64). For this work, we
do not consider UltraRAM, which is a different class of memory blocks on the
UltraScale+ architecture, but note that these can be exploited with the same
arguments as for BRAM (according to the principles in \secref{tiling}).
For benchmarked kernels we report the compute and memory utilization in terms of
the hardware constraints, with the primary bottleneck for I/O being BRAM, and
the bottleneck for performance varying between LUTs and DSPs, depending on the
data type. 

\subsection{Results}
\label{sec:results}

\begin{figure}
  \begin{minipage}{\columnwidth}
    \centering
    \includegraphics[width=\textwidth]{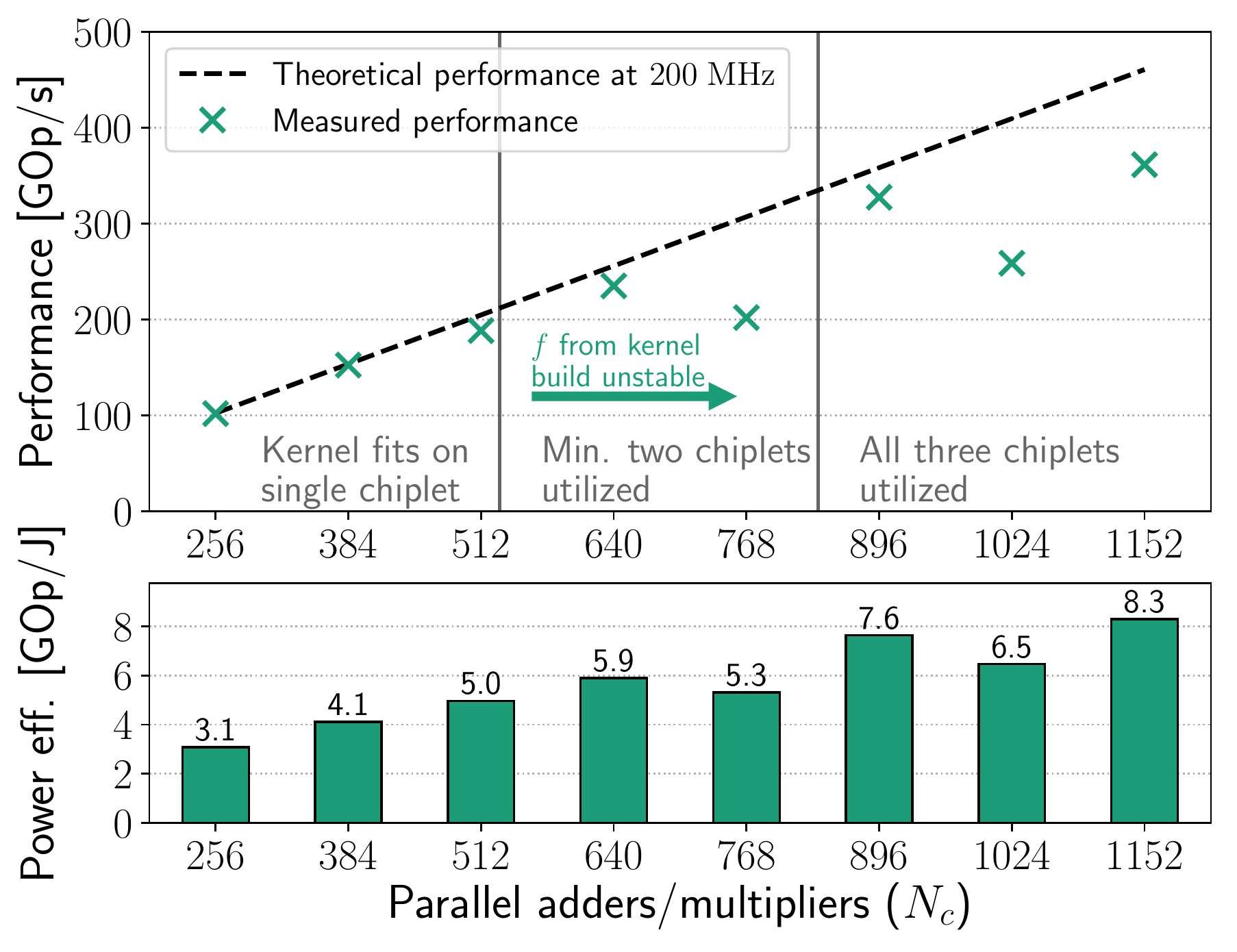}
    \vspace{-2em}
    \caption{Strong scaling for single precision floating point,
    $n{=}m{=}k{=}16384$ matrices.}
    \label{fig:perf_scaling}
  \end{minipage}
\end{figure}

We evaluate the computational performance and communication behavior of our
approach by constructing kernels within varying logic and storage budgets, based
on our C++ reference implementation.  To explore the scaling behavior with
increased parallelism, we measure strong scaling when increasing the number of
PEs, shown in \figref{perf_scaling}, by increasing $N_c$ for
$16384{\times}16384{\times}16384$ matrices. We report the median across 20 runs,
and omit confidence intervals, as all kernels behaved deterministically, making
errors negligible.  To measure power efficiency, we sample the direct current
power draw of the PSU in the host machine, then determine the FPGA power
consumption by computing the difference between the machine at idle with no FPGA
plugged in, and the FPGA plugged in while running the kernel. This method
includes power drawn by the full VCU1525 evaluation board, \emph{including the
integrated fan}. The kernels compile to maximum performance given by each
configurations at $\SI{200}{\mega\hertz}$ until the first chiplet/SLR crossing,
at which point the clock frequency starts degrading.  This indicates that the
chiplet crossings are the main contributor to long timing paths in the design
that bottleneck the frequency. 

\begin{table*}[h]
  \centering
  \resizebox{\textwidth}{!}{\begin{tabular}{l r r r r | r r r r r r r r}
    \textbf{Data type} & \textbf{$x_p$} & \textbf{$y_c$} & \textbf{$x_\text{tot}$} &
    \textbf{$y_\text{tot}$} & \textbf{Frequency} & \textbf{Performance} &
    \textbf{Power eff.} & \textbf{Arithm. int.} & \textbf{LUTs} & \textbf{FFs} & \textbf{DSPs} & \textbf{BRAM} \\\hline
    FP16 & $112$ & $16$ & $1904$ & $1920$ & $\SI{171.3}{\mega\hertz}$ &
    $\SI{606}{\giga\op\per\second}$ & $\SI{15.1}{\giga\op\per\joule}$ &
    $\SI{956}{\op\per\Byte}$ & $53\%$ & $24\%$ & $70\%$ & $90\%$ \\ 
    FP32 & $192$ & $8$ & $960$ & $1632$ & $\SI{145.7}{\mega\hertz}$ &
    $\SI{409}{\giga\op\per\second}$ & $\SI{10.9}{\giga\op\per\joule}$  &
    $\SI{302}{\op\per\Byte}$ & $81\%$ & $46\%$ & $48\%$ & $80\%$ \\ 
    FP64 & $96$ & $4$ & $864$ & $864$ & $\SI{181.2}{\mega\hertz}$ &
    $\SI{132}{\giga\op\per\second}$ & $\SI{3.13}{\giga\op\per\joule}$ &
    $\SI{108}{\op\per\Byte}$ & $38\%$ & $28\%$ & $80\%$ & $82\%$ \\
    uint8 & $132$ & $32$ & $1980$ & $2176$ & $\SI{186.5}{\mega\hertz}$ &
    $\SI{1544}{\giga\op\per\second}$ & $\SI{48.0}{\giga\op\per\joule}$  &
    $\SI{2073}{\op\per\Byte}$ & $15\%$ & $8\%$ & $83\%$ & $51\%$ \\ 
    uint16 & $210$ & $16$ & $1680$ & $2048$ & $\SI{190.0}{\mega\hertz}$ &
    $\SI{1217}{\giga\op\per\second}$ & $\SI{33.1}{\giga\op\per\joule}$ &
    $\SI{923}{\op\per\Byte}$ & $20\%$ & $11\%$ & $69\%$ & $88\%$ \\ 
    uint32 & $202$ & $8$ & $1212$ & $1360$ & $\SI{160.6}{\mega\hertz}$ &
    $\SI{505}{\giga\op\per\second}$ & $\SI{13.8}{\giga\op\per\joule}$ &
    $\SI{320}{\op\per\Byte}$ & $58\%$ & $11\%$ & $84\%$ & $86\%$ \\ 
  \end{tabular}}
  \vspace{1em}
  \caption{Highest performing kernels built for each data type.}
  \vspace{-2em}
  \label{tab:perf}
\end{table*}

\tabref{perf} shows the configuration parameters and measured results for the
highest performing kernel built using our architecture for half, single and
double precision floating point types, as well as 8-bit, 16-bit, and 32-bit
unsigned integer types. Timing issues from placement and routing are the main
bottleneck for all kernels, as the frequency for the final routed designs start
to be unstable beyond $33\%$ resource usage, when the number of chiplet
crossings becomes significant (shown in \figref{perf_scaling}). When resource
usage exceeds $80{-}90\%$, kernels fail to route or meet timing entirely.
Due to the large step size in BRAM consumption for large compute tiles when
targeting peak performance (see \secref{tiling}), some kernels consume less BRAM
than what would otherwise be feasible to route, as increasing the memory tile by
another stage of $N_{b,\text{min}}$ would exceed $N_{b,\text{max}}$. 

\begin{figure}
  \begin{minipage}{\columnwidth}
    \centering
    \includegraphics[width=.9\textwidth]{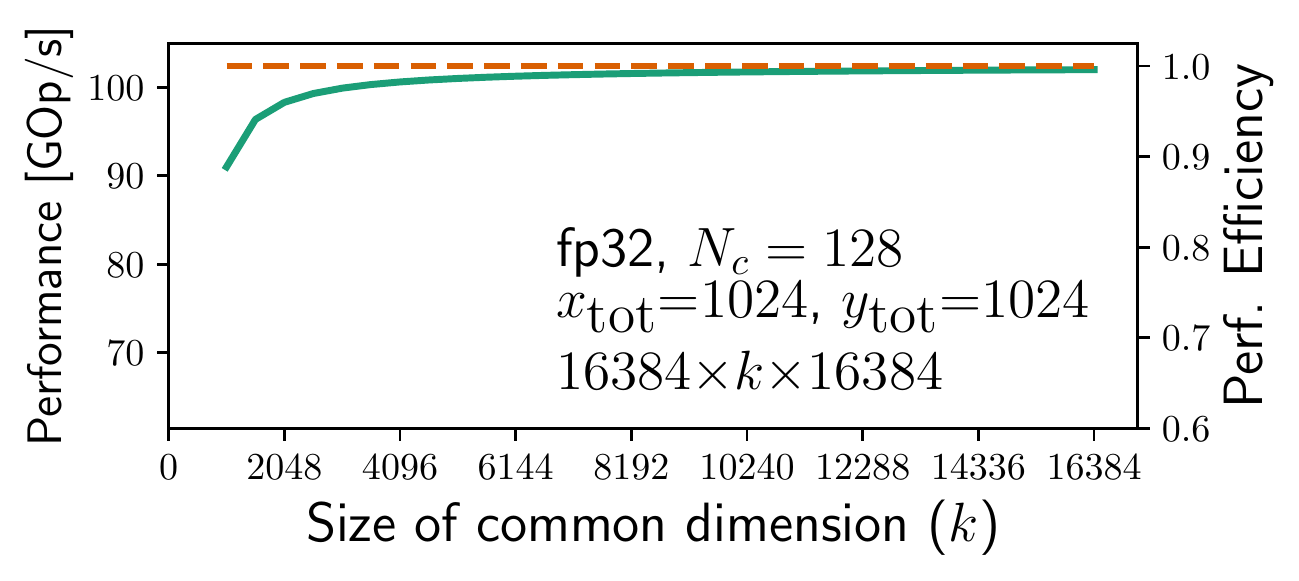}
  \end{minipage}
  \begin{minipage}{\columnwidth}
    \centering
    \includegraphics[width=.9\textwidth]{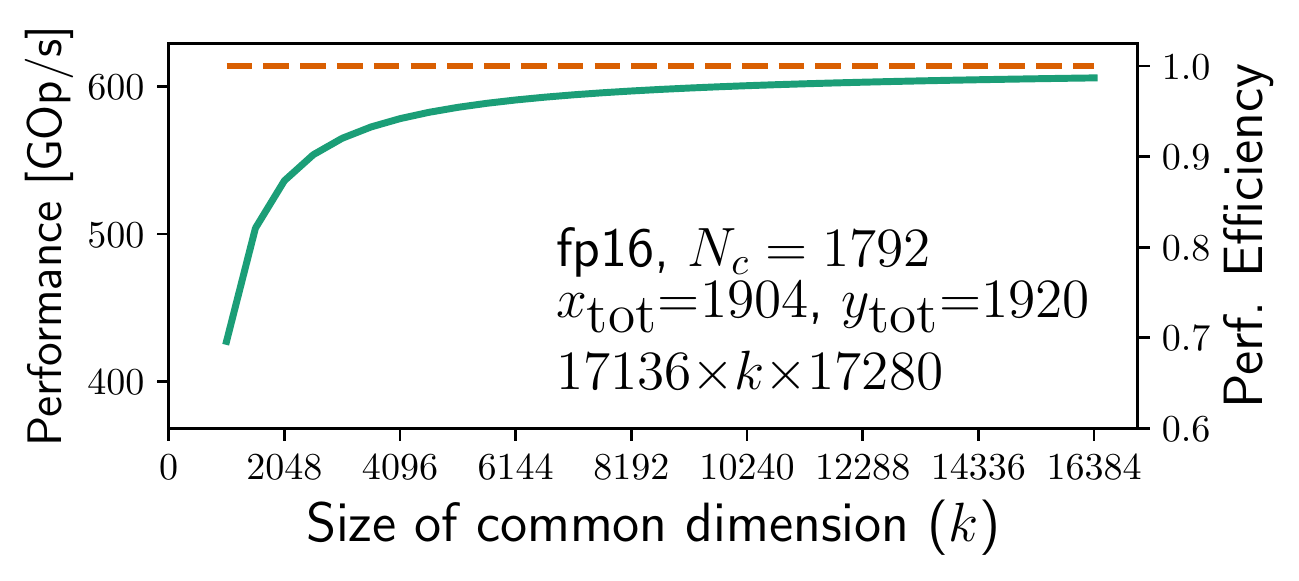}
  \end{minipage}
  \vspace{-1em}
  \caption{Fraction of maximum compute throughput for varying matrix size.}
  \vspace{-1em}
  \label{fig:efficiency}
\end{figure}

In contrast to previous implementations, we achieve optimal usage of the on-chip
memory by separating the drain phase of writing out results from the compute
phase. This requires the number of computations performed per memory tile to be
significantly larger than the number of cycles taken to write the tile out to
memory (see \secref{writing_back_results}). This effect is shown in
\figref{efficiency} for small $N_c$ (left) and large $N_c$ (right). For large
$N_c$, the time spent in draining the result is significant for small matrices.
In either scenario, optimal computational efficiency is approached for large
matrices, when there is sufficient work to do between draining each result tile.

\begin{figure}
  \centering
  \includegraphics[width=\columnwidth]{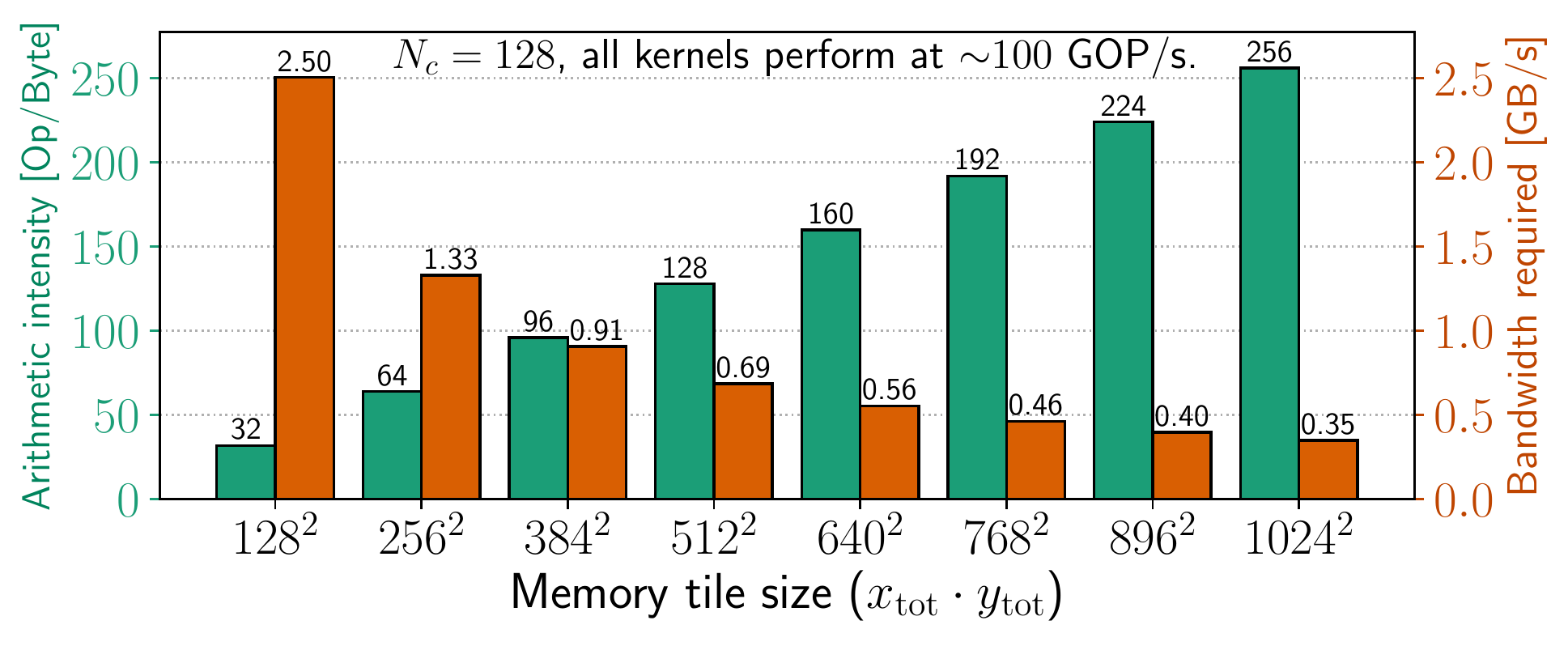}
  \vspace{-2em}
  \caption{FP32 arithmetic intensity with memory tile size.}
  \vspace{-1em}
  \label{fig:io_scaling}
\end{figure}

\figref{io_scaling} demonstrates the reduction in communication volume with
increasing values of the outer I/O tiles (i.e., $x_t x_b \cdot y_t y_b$). We
plot the arithmetic intensity, corresponding to $2{\times}$ the computational
intensity in \equref{memory_opt} (1 addition and 1 multiplication), and verify
that the communication volume reported by the runtime is verified to match the
analytical value computed with \equref{comm_vol}. We also report the average
bandwidth requirement needed to run each kernel (in practice, the bandwidth
consumption is not constant during runtime, as memory accesses are done as
bursts each time the row and column for a new outer product is loaded).  There
is a slight performance benefit from increasing memory tile size, as larger
tiles increase the ratio of cycles spent in the compute phase to cycles spent
writing back results, approaching perfect compute/DSP efficiency for large
matrices. 
For the largest tile size, the kernel consumes
$\SI{350}{\mega\byte\per\second}$ at $\SI{100}{\giga\op\per\second}$, which
corresponds to $\frac{350}{19200} = 1.8\%$ of the maximum bandwidth of a single
DDR4 module.  Even at the highest measured single precision performance
(\tabref{perf}) of $\SI{409}{\giga\op\per\second}$, the kernel requires
$\SI{1.35}{\giga\byte\per\second}$. This brings the I/O of matrix
multiplication down to a level where nearly the full bandwidth is left available. 

\vspace{0.35em}
\section{Related Work}
\label{sec:related_work}

Much of previous work focuses on the low level implementation for
performance~\cite{jovanovic2012fpga}, explores high-level
optimizations~\cite{gemm_dhollander}, or implements MMM in the context of neural
networks~\cite{gemm_dnn, gemm_intel}. 
To the best of our knowledge, this is the first work to minimize I/O of matrix
multiplication on FPGA \emph{in terms of hardware constants}, and the first work
to open source our implementation to benefit of the community. We relate this
paper to the most relevant works below. 

\tabref{comparison} shows a hybrid qualitative/quantitative comparison to
previously published MMM implementations on FPGA. Cells are left empty when
numbers are not reported by the authors, or when the given operation is not
supported. As our work is the only open source implementation, we are unable to
execute kernels from other works on the same FPGA, and resort to comparing the
performance reported in papers for the respective benchmarked FPGA. These FPGAs
thus vary widely in vendor, technology and architecture.

\begin{table*}
	\centering
	\sf
	\resizebox{\textwidth}{!}{\begin{tabular}{lrrrrrrrrrrrr}
		\toprule
	    & \textbf{Year}
      & \textbf{Device}
      & \makecell{\textbf{\% Logic}\\\textbf{util.}}
      & \makecell{\textbf{Freq.} \\ \textbf{$[\si{\mega\hertz}]$}}
	    & \makecell{\textbf{Perf. FP16}\\\textbf{[$\si{\giga\op\per\second}$]}} 
	    & \makecell{\textbf{Perf. FP32}\\\textbf{[$\si{\giga\op\per\second}$]}} 
	    & \makecell{\textbf{Perf. FP64}\\\textbf{[$\si{\giga\op\per\second}$]}} 
	    & \makecell{\textbf{Energy eff.} \\\textbf{FP32
      [$\si{\giga\op\per\joule}$]}}
      & \makecell{\textbf{Multiple}\\\textbf{data types}}
	    & \textbf{Lang.~(Portable)} 
	    & \makecell{\textbf{Open}\\\textbf{source}}
      & \makecell{\textbf{I/O}\\\textbf{model}} \\
		\midrule
     Zhuo~\cite{gemm_algorithms} & 2004 & Virtex-II~Pro & $98$ & $128$ & - & $2$
     & $2$ & - & \thumbsdown & HDL (\thumbsdown) & \thumbsdown & \thumbsdown \\ 
     Dou~\cite{gemm_systolic} & 2005 & Virtex-II~Pro & $99$ & $177$ & - & - &
     $39$ & - & \thumbsdown & HDL (\thumbsdown) & \thumbsdown & \thumbsdown \\ 
     Kumar~\cite{gemm_2009} & 2009 & Virtex-5 & $61$ & $373^\text{\textdagger}$
     & - & - & 30$^\text{\textdagger}$ & - & \thumbsdown & HDL (\thumbsdown) &
     \thumbsdown & \thumbsup \\ 
     Jovanovi\'c~\cite{jovanovic2012fpga} & 2012 & Virtex-6 & $100$ & $403$ & -
     & $203$ & - & - & \thumbsdown & HDL (\thumbsdown) & \thumbsdown &
     \thumbsdown \\ 
     D'Hollander~\cite{gemm_dhollander} & 2016 & Zynq-7000 & $99$ & $100$ & - &
     $5$ & - & - & \thumbsdown & HLS (\thumbsup) & \thumbsdown & 
      \thumbsdown \\ 
		 Guan~\cite{gemm_dnn} & 2017 & Stratix V & $95$ & $150$ & - & $100$ & - & $2.92$ & \thumbsup & HDL/HLS (\thumbsdown) & \thumbsdown & \thumbsdown \\
     Moss~\cite{gemm_intel} & 2018 & HARPv2 & $99$ & $313$ & - & $800$ & - &
     $22.0$ & \thumbsup & HDL (\thumbsdown) & \thumbsdown & 
     \thumbsdown \\ 
     This~work & 2019 & VCU1525 & $69{-}90$ & $146{-}190$ & $606$ & $409$ &
     $122$ & $10.9$ & \thumbsup & HLS (\thumbsup) & \thumbsup & \thumbsup \\ 
		\bottomrule
	\end{tabular}}
  \caption{Comparison to previous FPGA implementations.
  $^\text{\textdagger}$Simulation only.}
	\label{tab:comparison}
\end{table*}

Zhuo and Prasanna~\cite{gemm_algorithms} discuss two matrix multiplication
implementations on FPGA, and include routing in their considerations, and
support multiple floating point precisions. The authors suggest two algorithms,
where both require a number of PEs proportional to the matrix size. While these
only require loading each matrix once, they do not support matrices of
arbitrary size, and thus do not scale without additional CPU orchestration. 

Dou~et~al.~\cite{gemm_systolic} design a linear array of processing elements,
implementing 64-bit floating point matrix multiplication -- no support is
offered for other data types, as the work emphasizes the low-level
implementation of the floating point units. The authors derive the required
off-chip bandwidth and buffer space required to achieve peak performance on the
target device, but do not model or optimize I/O in terms of their buffer space
usage, and do not report their tile sizes or how they were chosen. Furthermore,
the authors double-buffer the output tile, reducing the maximum achievable
computational intensity by a factor $\sqrt{2}$ (see
\secref{writing_back_results}).

A customizable matrix multiplication implementation for deep neural network
applications on the Intel HARPv2 hybrid CPU/FPGA platform is presented by
Moss~et~al.~\cite{gemm_intel}, targeting single precision floating point (FP32),
and fixed point/integer types. The authors exploit native floating point DSPs on
an Arria~10 device to perform accumulation, and do not consider data types that
cannot be natively accumulated on their chip, such as half or double precision.
The I/O characteristics of the approach is not reported quantitatively.
Wu~et~al.~\cite{xilinx_dnn_frequency} present a highly specialized architecture
for maximizing DSP usage and frequency of $\SI{16}{\bit}$ integer matrix
multiplication for DNN acceleration on two Xilinx UltraScale chips, showing how
peak DSP utilization and frequency can be reached, at the expense of generality,
as the approach relies on low-level details of the chips' architecture, and as
no other data types are supported. 

Kumar~et~al.~\cite{gemm_2009} provide an analysis of the trade-off between I/O
bandwidth and on-chip memory for their implementation of 64-bit matrix
multiplication. The authors arrive at a square output tile when deriving the
constraints for overlapping I/O, although the derived computational intensity is
reduced by a factor $\sqrt{2}$ as above from double buffering. In our model, the
fast memory utilization is captured explicitly, and is maximized in terms of
on-chip memory characteristics of the target FPGA, allowing tile sizes that
optimize both computational performance and computational intensity to be
derived directly. Lin~and~Leong~\cite{leong} model sparse MMM, with dense MMM as
a special case, and project that even dense matrix multiplication may become I/O
bound in future FPGA generations. Our model guides how to maximize utilization
in terms of available on-chip memory to mitigate this, by capturing their
characteristics in the tiling hierarchy.

Finally, the works above were implemented in hardware description languages, and
\emph{do not disclose the source code allowing their findings to be reproduced
or ported to other FPGAs}. For the results presented here, we provide a
high-level open source implementation, to encourage reusability and portability
of FPGA codes.

Designing I/O minimizing algorithms has been an active field of research for
more than 40 years. Starting with register allocation
problems~\cite{registerpebblecolor}, through single processor, two-level memory
system~\cite{red_blue_pebble_game}, distributed systems with fixed
~\cite{toledo} and variable memory size~\cite{25d}. Although most of the work
focus on linear algebra~\cite{25d, toledo, Cholesky} due to its regular access
pattern and powerful techniques like polyhedral modeling, the implication of
these optimizations far exceeds this domain. Gholami et
al.~\cite{dnnParallelism} studied model and data parallelism of DNN in the
context of minimizing I/O for matrix multiplication routines. Demmel and
Dinh~\cite{comm_DNN} analyzed I/O optimal tiling strategies for convolutional
layers of NN. 
 
\section{Conclusion}
\label{sec:conclusion}

We present a high-performance, open source, flexible, portable, and scalable
matrix-matrix multiplication implementation on FPGA, which simultaneously
maximizes performance and minimizes off-chip data movement.  By starting from a
general model for computation, I/O, and resource consumption, we create a
hardware architecture that is optimized to the resources available on a target
device, and is thus not tied to specific hardware.
We evaluate our implementation on a wide variety of data types and
configurations, showing $\SI{409}{\giga\op\per\second}$ 32-bit floating point
performance, and $\SI{1.5}{\tera\op\per\second}$ 8-bit integer performance,
utilizing ${>}80\%$ of hardware resources. We show that our model-driven I/O
optimal design is robust and high-performant in practice, yielding better or
comparable performance to HDL-based implementations, and conserving bandwidth to
off-chip memory, while being easy to configure, maintain and modify through the
high-level HLS source code.

\section*{Acknowledgments}

We thank Xilinx for generous donations of software and hardware.  This project
has received funding from the European Research Council (ERC) under the European
Union's Horizon 2020 programme (grant agreement DAPP, No. 678880).

\bibliographystyle{ACM-Reference-Format}
\bibliography{GEMM_FPGA}

\end{document}